\documentclass[12pt]{article}
\usepackage{amsmath,amssymb}
\def\R{\hbox{{\rm I}\kern-0.2em{\rm R}\kern0.2em}}%definition of reals
\def\R{\hbox{{\rm I}\kern-0.2em{\rm R}\kern0.2em}}%mathematical R for reals
\def\D{\hbox{{\rm I}\kern-0.2em{\rm D}\kern0.2em}}

\def\be{\begin{equation}}
\def\ee{\end{equation}}

\def\({\left(}
\def\){\right)}
\def\[{\left[}
\def\]{\right]}
\def\bc{\begin{center}}
\def\ec{\end{center}}

\begin{document}

{\large \bf APPROXIMATE SYMMETRIES OF LAGRANGIANS FOR PLANE
SYMMETRIC GRAVITATIONAL WAVE-LIKE SPACETIMES}

\textit{IBRAR HUSSAIN} and \textit{ASGHAR QADIR}

$^{\dag}$Centre for Advanced Mathematics and Physics\\
National University of Sciences and Technology\\
Campus of the College of Electrical and Mechanical Engineering\\
Peshawar Road, Rawalpindi, Pakistan

E-mail: ihussain@camp.edu.pk, aqadirmath@yahoo.com

{\bf Abstract}. Using Lie symmetry methods for differential
equations we have investigated the symmetries of a Lagrangian for a
plane symmetric static spacetime. Perturbing this Lagrangian we
explore its approximate symmetries. It has a non-trivial first-order
approximate symmetry.

\section{Introduction}

The problem of defining energy (or mass) in general relativity
arises from the fact that arbitrary spacetimes can be non-static (or
even non-stationary) and hence global (or even local) energy
conservation may be lost. For static spacetimes there exists a
timelike \textit{isometry} or \textit{Killing Vector} (KV)
\cite{ESEFEs}, which can be used to define the energy, $E$, of a
test-particle by $E={\bf k}\cdot{\bf p}$, where ${\bf k}$ is the
timelike KV  and ${\bf p}$ is the momentum 4-vector of the
test-particle. Further, energy conservation in the spacetime is
guaranteed in the frame using ${\bf k}$ to define the time
direction. However if there does not exist a timelike KV, energy is
not conserved and hence the energy of a test particle cannot be
defined. Since gravitational waves {\it must} be given by non-static
spacetimes, the problem of defining the energy content of
gravitational waves is particularly severe.

Minkowski spacetime is maximally symmetric having 10 KVs which form
the Poincar\'{e}, algebra $so(1,3)\oplus _s \R^{4}$, where $\oplus
_s$ denotes a semi direct sum (the algebra is denoted by
$so(1,3)\oplus _s \R^{4}$ while the group is denoted by
$SO(1,3)\oplus _s \R^{4}$) \cite{ESEFEs}. The generators of this
algebra give conservation laws for energy, spin angular momentum and
linear momentum \cite{AQ}. Going to the Schwarzschild and
Reissnor-Nordstr\"{o}m spacetimes we lose linear and spin angular
momentum conservation while in the Kerr spacetime we lose two more
conservation laws \cite{ESEFEs}. Approximate symmetries for
differential equations (DEs) have been used to recover these lost
conservation laws \cite{KMQ,IMQ}.

Different people have tried to define ``approximate symmetry''
\cite{a21,a22,a23,a24,a25}, in which the broken time-translation
symmetry provides information about the energy content of the
gravitational wave spacetimes, but none of them have been
unequivocally successful. Nevertheless, this approach of a {\it
slightly} broken symmetry seems promising.

It was proposed \cite{IMQ} that the concept of ``approximate
symmetry'' of DEs \cite{Ib} could be extended and adapted for the
purpose of defining energy in gravitational waves by using the
connection found between isometries and symmetries of DEs through
the geodesic equations \cite{a13,a19}. In this paper we will use
this concept of approximate symmetries of DEs to look for a
resolution of the problem of energy in gravitational wave
spacetimes.

The plan of the paper is as follows. In the next section we briefly
review the definitions of symmetries and approximate symmetries of a
Lagrangian. In section 3 symmetries of a Lagrangian for plane
symmetric static spacetime and approximate symmetries of a perturbed
Lagrangian for a plane symmetric gravitational waves like spacetime
are discussed. Finally a summary and discussion is given in section
4.

\section{Symmetries and approximate symmetries of a Lagrangian}

Noether's theorem \cite{Nth} relates the constants of motion of a
given Lagrangian system to its symmetry transformation \cite{Ib,
KM}. Symmetry generators of a Lagrangian of a manifold form a Lie
algebra \cite{BKK} and from the geometric point of view symmetries
of a manifold are characterized by its KVs, which always form a
finite dimensional Lie algebra \cite{HkEl}.

In general a manifold may not possess an exact symmetry but may
approximately do so. It would be of interest to look at the
approximate symmetries of the manifold. They form an approximate Lie
algebra \cite{Gaz}. Methods for obtaining exact and approximate
symmetries of a Lagrangian are available in the literature
\cite{Ib,WF,TK}.

Symmetries of a Lagrangian, also known as Noether symmetries, are
defined as follows. Consider a vector field \cite{Ib}
\begin{equation}
\mathbf X=\xi (s,x^i) \partial/\partial s+\eta^j(s,x^i)
\partial/\partial x^j \quad (i,j=0,...,3). \label{1}
\end{equation}
Its first prolongation is
\begin{equation}
\mathbf{X}^{[1]}=\mathbf{X}+(\eta^j_{,s}+\eta^j_{,i}x^{i \prime}
-\xi_{,s}x^{j \prime}-\xi_{,i}x^{i \prime}x^{j \prime})
\partial/\partial x^{j \prime},\label{2}
\end{equation}
where $``\prime"$ denote differentiation with respect to $s$. Now
consider a set of second-order ordinary differential equations
\begin{equation}
x^{i {\prime \prime}}=g(s,x^i,x^{i \prime}), \label{3}
\end{equation}
which has a Lagrangian $L(s,x^i,x^{i \prime})$. Then {\bf X} is a
Noether point symmetry of the Lagrangian $L(s,x^i,x^{i \prime})$ if
there exists a function $A(s,x^i)$ such that
\begin{equation}
\mathbf{X}^{[1]}L+(D_s \xi)L=D_{s}A,  \label{4}
\end{equation}
where
\begin{equation}
D_s=\partial/\partial s+x^{i \prime} \partial/\partial x^i.
\label{5}
\end{equation}
For more general considerations see \cite{Ib}.

First-order approximate symmetries of the Lagrangian are defined as
follows \cite{TK}. For a first-order perturbed system of equations,
$\mathbf{E}=\mathbf{E}_0+\epsilon\mathbf{E}_1=O(\epsilon ^2)$
corresponding to the first-order perturbed Lagrangian, $L(s,x^i,x^{i
\prime},\epsilon)=L_0(s,x^i,x^{i \prime})+ {\epsilon}L_1(s,x^i,x^{i
\prime} )+O(\epsilon ^2)$, the functional $\int_V Lds$ is invariant
under the one parameter group of transformations with {\it
approximate Lie symmetry generator}
$\mathbf{X}=\mathbf{X}_0+\epsilon\mathbf{X}_1+O(\epsilon ^2)$ up to
the gauge $A=A_0+\epsilon A_1$, where $\mathbf{X}_0^{[1]}L_0+(D_s
\xi_0)L_0=D_sA_0$ and $\mathbf{X}_1^{[1]}L_0+\mathbf{X}_0^{[1]}L_1
+(D_s \xi_1)L_0+(D_s \xi_0)L_1=D_sA_1$. Here $\mathbf{X}_0$ is the
exact symmetry generator and $\mathbf{X}_1$ the {\it first-order
approximate symmetry} generator. The perturbed equations always have
the trivial approximate symmetry generator $\epsilon \mathbf{X}_0$
and $\mathbf{X}$ with $\mathbf{X}_0\neq 0 \neq \mathbf{X}_1\neq
k\mathbf{X}_0$, is called a {\it non-trivial} approximate symmetry.

\section{Symmetries and approximate symmetries of a Lagrangian for
a plane symmetric gravitational wave-like spacetime}

To try to find a resolution of the problem of definition of energy
in gravitational wave spacetimes, we consider a static spacetime and
then perturb it time dependently (for definiteness by $\epsilon$$t$)
to make it slightly non-static. For this purpose we take a plane
symmetric static metric \cite{TQZ}
\begin{equation}
ds^{2}= e^{2\nu(x)}dt^2-dx^2-e^{2\mu(x)}(dy^2+dz^2), \label{6}
\end{equation}
with $\mu(x)=\nu^2(x)=(x/X)^2$, where $X$ is a constant having the
same dimensions as $x$.

The Lagrangian defined by minimizing the arc length in (6) is
\begin{equation}
L= e^{2x/X}\dot{t}^2-\dot{x}^2-e^{2x^2/X^2}(\dot{y}^2+\dot{z}^2),
\label{7}
\end{equation}
where ``$\cdot$" denotes differentiation with respect to the
geodetic parameter $s$. Using (7) in (4) we obtain a set of 19
determining PDEs for 6 unknown functions $\xi$, $\eta_{i}$
($i=0,...,3$) and $A$, where each of these is a function of the 5
variables $s$ and $x^i$. Solving these equations we get the symmetry
generators
\begin{equation}
\quad \mathbf{X}_{0}=\frac{\partial}{\partial t},\quad
\mathbf{X}_{1}=\frac{\partial}{\partial y},\quad
\mathbf{X}_{2}=\frac{\partial}{\partial z},\quad
\mathbf{X}_{3}=y\frac{\partial}{\partial
z}-z\frac{\partial}{\partial y},\quad
\mathbf{Y}_{0}=\frac{\partial}{\partial s},\quad A=c , \label{8}
\end{equation}
where $c$ is a constant, $\mathbf{X}_{0}$ corresponds to energy
conservation, $\mathbf{X}_{1}$ and $\mathbf{X}_{2}$ correspond to
linear momentum conservation along $y$ and $z$, while
$\mathbf{X}_{3}$ corresponds to angular momentum conservation in the
$yz$ plane.

For the approximate symmetries of a Lagrangian for this plane
symmetric gravitational wave-like spacetime we consider
$\nu(x)=2(x/X+\epsilon t/T)$ and $\mu(x)=2(x^2/X^2+\epsilon t/T)$ in
the metric (6), where $T$ is a constant having dimensions of $t$.
Its first-order perturbed Lagrangian is
\begin{equation}
\quad
L=e^{2x/X}\dot{t}^2-\dot{x}^2-e^{2x^2/X^2}(\dot{y}^2+\dot{z}^2)+
2\epsilon t[e^{2x/X}\dot{t}^2-e^{2x^2/X^2}(\dot{y}^2+\dot{z}^2)]/T
+O(\epsilon^2). \label{9} \
\end{equation}
Using the exact symmetry generators given by (8) and solving the
system of determining equations for approximate symmetries, we get
the non-trivial approximate symmetry $\mathbf{X}_a$, along with the
trivial symmetries, and the gauge function $A_{1}$ is again a
constant,
\begin{equation}
\mathbf{X}_a= \partial/\partial t-\epsilon (t \partial/\partial t +
y \partial/\partial y +z \partial/\partial z)/T. \label{10}
\end{equation}
The physical meaning of this non-trivial approximate symmetry found
here is worth exploring.

\section{Summary and Discussion}
We addressed the problem of energy in gravitational wave spacetimes.
For this purpose we first considered a plane symmetric static
spacetime that has $4$ KVs \cite{TQZ}. The Lagrangian for this
metric has an additional symmetry $\partial/\partial s$. Since
gravitational waves must be given by non-static metrics, we
perturbed the static metric here with a term $\epsilon t$ and
retained the terms containing $\epsilon$, neglecting its higher
powers. For the Lagrangian of this perturbed metric we found a
non-trivial first-order approximate symmetry given by (10). The
trivial first-order approximate timelike Noether symmetry, for the
perturbed Lagrangian (9), can be interpreted to give the extent of
energy non-conservation, and hence (possibly) the energy content, in
the gravitational wave-like spacetime. Since the stress-energy
tensor is non-zero \cite{TQZ}, we need to consider the exact
solution to understand how much energy is in the field and how much
in its interaction with matter.

\section*{Acknowledgments} We thank NUST for support and Salento
University for hospitality.

%\section*{ References }


\begin{thebibliography}{99}
\bibitem{ESEFEs} Kramer D, Stephani H, MacCullum, M. A. H., and Herlt E, \textit{Exact
Solutions of Einstein Field Equations}, Cambridge University Press,
Cambridge, 1980.

\bibitem{AQ} Qadir A, {\it Applications of Symmetry Methods}, pp.45-73, eds.
Qadir A and Saifullah K, National Centre for Physics, Islamabad,
2006.

\bibitem{KMQ} Kara, A. H., Mahomed, F. M., and Qadir A, ``Approximate
Symmetries and Conservation Laws of the Geodesic Equations for the
Schwarzschild Metric", \textit{Nonlinear Dynamics} (to appear).

\bibitem{IMQ} Hussain, I. Mahomed, F. M., and Qadir A, \textit{SIGMA}, \textbf{3}
(2007) 9 pages, arXiv:0712.1089.

\bibitem{a21} Komar A, \textit{Phys. Rev}., \textbf{127} (1962) 1411;
\textbf{129} (1963) 1873.

\bibitem{a22} Matzner R, \textit{J. Math. Phys}., \textbf{9} (1968) 1063;
\textbf{10} (1968) 1657.

\bibitem{a23} Isaacson, R. A., \textit{Phys. Rev}., \textbf{166} (1968)
1263; 1272.

\bibitem{a24} York, J. W., \textit{Ann. Inst. Henri Poincar$\grave{e}$ XXI,}
\textbf{4} (1974) 319.

\bibitem{a25} Spero A, and Baierlein R, \textit{J. Math. Phys}., \textbf{18}
(1977) 1330; \textbf{19} (1978) 1342.

\bibitem{Ib} Ibragimov, N. H., \textit{Elementary Lie group Analysis and
Ordinary Differential Equations}, Wiely, Chichester, 1999.

\bibitem{a13} Feroze T, Mahomed, F. M., and Qadir A, \textit{Nonlinear
Dynamics}, \textbf{45 }(2006) 65.\textit{\ }

\bibitem{a19} Aminova A. V., \textit{Sobrink Mathematics}, \textbf{186},
(1995) 1711.

\bibitem{Nth} Noether E, "Invariant variations problems", \textit{Nachr. Konig.
Gissell. Wissen., Gottingen, Math.-Phys.Kl}. \textbf{2} (1918) 235.
(English translation in transport theory and Statistical Physics
\textbf{1} (1971)) 186.

\bibitem{KM} Kara, A. H., Mahomed, F. M., \textit{J. Nonlinear Math. Phys}.,
\textbf{9} (2002) 60.

\bibitem{BKK} Bokhari, A.H., Kara, A. H., Kashif, A. R., and Zaman, F. D.,
\textit{International. J. Theoretical. Phys}., \textbf{45} (2006)
1063.

\bibitem{HkEl} Hawking, S. W., and Ellis, G. F. R., \textit{The Large Scale Structure
of Spacetime}, Cambridge University Press, Cambridge, 1973.

\bibitem{Gaz} Gazizov, R. K., \textit{J. Nonlinear Math. Phys}., \textbf{3} (1996) 96.

\bibitem{WF} Wafo Soh. C., and Mahomed, F. M., \textit{Class. Quantum Grav}.
\textbf{16} (1999) 3553.

\bibitem{TK} Feroze T. and Kara, A.H., \textit{Inernational. J. Non-linear
Mechanics} \textbf{37} (2002) 275.

\bibitem{TQZ} Feroze T, Qadir A, and Ziad M, \textit{J. Math.
Phys}., \textbf{42} (2001) 4947.

\end{thebibliography}
\end{document}